\documentclass[review]{elsarticle}
\usepackage{lineno,hyperref}
\usepackage{times}
\usepackage{balance}
\usepackage{amsmath}
\usepackage[tight,footnotesize]{subfigure}
\usepackage{graphicx}
\usepackage{bm}
\usepackage{algorithm}
\usepackage[noend]{algorithmic}
\usepackage{multirow}
\usepackage{slashbox}
\usepackage{caption}
\usepackage{amsfonts}
\newtheorem{theorem}{Theorem}

\newtheorem{lemma}{Lemma}

\modulolinenumbers[5]

\journal{Journal of \LaTeX\ Templates}

\begin{document}

\begin{frontmatter}

\title{Product Sequencing and Pricing under Cascade Browse Model}

\author{Shaojie Tang}
\address{Naveen Jindal School of Management, University of Texas at Dallas}

\author{Jing Yuan}
\address{Department of Computer Science, University of Texas at Dallas}
%


\begin{abstract}
In this paper, we study the joint product sequencing and pricing problem faced by many online retailers such as Amazon. We assume that a consumer's purchasing behavior can be explained by a ``consider-then-choose'' model: she first forms a consideration set by screening  a subset of products sequentially, and then decides which product to purchase from her consideration set. We propose a \emph{cascade browse model} to capture the consumer's browsing behavior, and use the Multinomial Logit (MNL) model as our choice model. We study two problems in this paper: in the first problem, we assume that each product has a fixed revenue and preference weight, the goal is to identify the best sequence of products to offer so as to maximize the expected revenue subject to a cardinality constraint. We propose a constant approximate solution to this problem. As a byproduct, we propose the first fully  polynomial-time approximation scheme  (FPTAS) for the classic assortment optimization problem subject to one capacity constraint and one cardinality constraint. In the second problem, we treat the price of each product as  a decision variable and our objective is to jointly decide a sequence of product and their prices to maximize the expected revenue. We propose a constant approximate solution to this problem.
\end{abstract}


\end{frontmatter}

%

\section{Introduction}\label{sec:introduction}
 In this paper, we consider the setting where the platform has a set of products and a limited number of  vertically differentiated display positions. Our objective is to select a sequence of  products, as well as their prices, so as to maximize the expected revenue. We assume that the consumer's purchasing behavior is governed by a ``consider-then-choose'' model.  Once a sequence of products is displayed to a consumer, she first forms her consideration set by browsing the products sequentially, then chooses one product (or nothing) from her consideration set based on MNL model. This modeling approach is well established in the literature in quantitative marketing and operations  \cite{aouad2019click}.

We depart from existing  literature in assuming that the browsing behavior of a consumer is governed by a cascade browse model. Our model is inspired by the cascade click model proposed in \cite{craswell2008experimental}, however, their purpose is different from ours, e.g., their goal is to find a good model that captures the click behavior of a consumer in the context of online advertising. We next give a brief introduction to our cascade browse model. Under the cascade browse model, each product is associated with a continuation probability which represents the probability that a consumer continues to browse the next product after browsing the current product. The purchasing decision process of a customer can be roughly described as follows. Upon viewing a sequence of products, she adds the product displayed in the first position to her consideration set. Then she will decide whether to make a purchasing decision, including non-purchase option, or to continue adding the product in the next position to her consideration set. Once a consumer decides to make a purchase from among her current consideration set, we assume that her purchasing decision is governed by MNL model, and she will leave the system after the purchase. Otherwise, if she decides to continue adding the next product to her consideration set, the whole process continues with the next product until she makes a purchasing decision at some point. 

Our main contributions are summarized as follows.
\begin{enumerate}
\item We introduce a cascade browse model to capture the consumer's browsing behavior. Our model not only captures the position-bias effect but also considers the externality among displayed products.
\item We propose constant approximate solutions to the corresponding assortment optimization and pricing problems.
 \item As a byproduct, we propose the first fully  polynomial-time approximation scheme  (FPTAS) for the classic assortment optimization problem subject to one capacity constraint and one cardinality      constraint.
\end{enumerate}

\textbf{Literature Review:} Our work is closely related to the assortment optimization problems \cite{li2015d,davis2014assortment,blanchet2016markov,farias2013nonparametric,aouad2015assortment}, which have been extensively studied in the revenue management literature. We have limited our review to studies under the MNL model with position bias. \cite{davis2013assortment} and \cite{abeliuk2016assortment} were the first to study the assortment optimization problem under MNL model with position bias. However, they assume that consumers browse all displayed products, and the location of a product only affects its  MNL-based preference weight. In our study, we assume that the consumer only browses a subset of  displayed products, and then makes a purchase from among all browsed products. Recently, \cite{aouad2015display,ferreira2019learning} considers the assortment optimization problem with  vertically differentiated locations under MNL model. Similar to our work, they adopt the consider-then-choose model, e.g., the consumer first browses a random number of products and
then makes a choice within these products according to the MNL model. Different from our cascade browse model, their model does not capture the externality among displayed products, e.g., they assume that the probability of a displayed product being browsed is solely dependent on its location, which is not affected by other displayed products. \cite{aouad2019click} also study the consider-then-choose model, however, they assume that each product is considered by a customer with a fixed probability. Another stream of literature introduces the search cost in the consumer choice model. For example, \cite{wang2017impact}  assume that a consumer selects a group of products as her consideration set such that the expected utility net of search cost is maximized. Our study is also related to assortment pricing. In addition to the literature on assortment pricing under standard MNL model \cite{anderson1992discrete,song2007demand,hopp2005product}, \cite{najafi2019multi} studied  a pricing problem based on a cascade click model. Our model differs from theirs in that we use a combination of cascade browse model and MNL model to capture the consumer's purchasing behavior. Moreover, they assume that the sequence of displayed products is fixed.

\section{Preliminaries and Problem Formulation}
\label{sec:system}
We consider the setting where the platform has $N$ products $\Omega=[N]$ and $B$ vertically differentiated display positions. Our objective is to allocate $B$ products to $B$ display locations so as to maximize the expected revenue of the platform. We use a ``consider-then-choose'' model to capture the consumers' purchasing behavior. Once products are allocated to display locations, a consumer chooses among products in two phases: she first forms a \emph{consideration set} by  sequentially examining the products according to the linear order of locations, then decides which product to purchase from among her consideration set. In this paper, we  propose a cascade browse model to model the consumer's browsing behavior in the first phase and use the MNL model to capture the consumer's purchase behavior in the second phase.   We next introduce our ``consider-then-choose'' model in more details.

\paragraph{Some notations}    Throughout this paper, we use capital letter to denote a \emph{sorted sequence} of products, and use corresponding calligraphy letter  to denote a \emph{set} of products.  For example, given a   sequence of products $S$, we use $\mathcal{S}$ denote the set of products involved in $S$. Moreover, we use $S_{\leq i}$ (resp. $S_{< i}$, $S_{> i}$, $S_{\geq i}$) to denote the longest subsequence of $S$ which is placed no later than (resp. before, after, no earlier than) product $i$. Correspondingly, we use $\mathcal{S}_{\leq i}$ (resp. $\mathcal{S}_{< i}$, $\mathcal{S}_{> i}$, $\mathcal{S}_{\geq i}$) to denote the set of products in $S_{\leq i}$ (resp. $S_{< i}$, $Q_{> i}$, $S_{\geq i}$). For example, given a sequence of prodicts $S=\{3,4,1,2,5\}$, $S_{< 1}=\{3,4\}$, we have $S_{> 1}=\{2,5\}$, $S_{\leq 1}=\{3,4,1\}$, and  $S_{\geq 1}=\{1,2,5\}$.  Given two sequences $S_1$ and $S_2$, we define $S_1 \oplus S_2$ as a new sequence by first displaying $S_1$ and then displaying $S_2$.
\subsection{Phase 1: Forming a consideration set}
Under our cascade browse model, each product $i\in \Omega$ has a continuation probability $\theta_i$ which represents the probability that a consumer continues to browse the next product (if any) after browsing $i$.

Given a sequence of products $S$ and a product $i\in S$, we define the \emph{reachability} $\Theta^S_{i}$  of $i$ given $S$ as the probability that a consumer browses $i$ given $S$:
\[\Theta^S_{i}= \prod_{j\in \mathcal{S}_{<i}}\theta_{j}\]

It follows that for any $S$ and $i\in S$, the probability $\Pr[\mathcal{S}_{\leq i}\mid S]$ that a consumer browses all and only products in $\mathcal{S}_{\leq i}$ is
$$
\Pr[\mathcal{S}_{\leq i}\mid S]=
\begin{cases}
\Theta^S_{i}, \mbox{if $i$ is the last product in $S$}\\
\Theta^S_{i}(1-\theta_i), \mbox{ otherwise}
\end{cases}
$$

We refer to $\mathcal{S}_{\leq i}$ as the random consideration set  induced by $S$.

\subsection{Phase 2: Making a purchase}
In the second phase, the consumer follows MNL model to make a purchase from among her consideration set. In the MNL model, each product $i\in \Omega$ is associated with a revenue $\alpha_i$, and a MNL-based preference weight $\beta_i$. For any given realized consideration set $\mathcal{S}$, the probability $P_i(\mathcal{S})$ that $i\in \mathcal{S}$ is purchased by a single representative consumer is
\begin{equation}
\label{eq:555}
P_i(\mathcal{S})=\frac{\beta_i}{\sum_{i\in \mathcal{S}}\beta_i+1}
\end{equation}

The expected revenue $g(\mathcal{S})$ of the consideration set $\mathcal{S}$ is
\begin{equation}
\label{eq:888}
g(\mathcal{S})=\sum_{i\in \mathcal{S}} \alpha_i P_i(\mathcal{S})
\end{equation}

The expected revenue $f(S)$ of a sequence of products $S$ can be calculated  as
\[f(S)=\sum_{i\in \mathcal{S}}\Pr[\mathcal{S}_{\leq i}\mid S] g(\mathcal{S}_{\leq i})\]

\subsection{Problem Formulation}
We study two problems in this paper. In the first problem, given that  $\alpha_i$ and $\beta_i$ are constants for each $i\in \Omega$, we aim to determine the best sequence of products that maximizes the expected revenue. In the second problem, we relax the assumption that $\alpha_i$ and $\beta_i$ are pre-fixed and assume that $\alpha_i$ and $\beta_i$ are functions of $i$'s price. By treating the price of each product as a decision variable, we study the joint product selection, sequencing, and pricing problem.
\subsubsection{Revenue maximizing product selection and sequencing}
We first introduce the revenue maximizing product selection and sequencing problem. Assume $\alpha_i$ and $\beta_i$ are given  for each $i\in \Omega$, our objective is to identify the best subset of $B$ products, as well as their sequence, so as to maximize the expected revenue. We  present the formal definition of our problem in $\textbf{P.A}$.
 \begin{center}
\framebox[0.45\textwidth][c]{
\enspace
\begin{minipage}[t]{0.45\textwidth}
\small
$\textbf{P.A}$
$\max f(S)$\\
\textbf{subject to:} $|S|\leq B$;
\end{minipage}
}
\end{center}
\vspace{0.1in}
\subsubsection{Joint product selection, sequencing, and pricing}
\label{sec:111}
In addition to optimizing the positioning of the products, the platform could also adjust the price of each product to improve the expected revenue.  We next study the case when the price of a product is also a decision variable. To capture the impact of price on a consumer's purchase decision, it is common practise to interpret  $\alpha_i$, as well as $\beta_i$, as a function of $i$'s quality, price, and cost  for any $i\in \Omega$.  Assume each product $i\in \Omega$ is associated with a \emph{fixed} quality $q_i$, an \emph{adjustable} price $p_i$, and a \emph{fixed} cost $c_i$. If we set $\alpha_i=p_i-c_i$ and  $\beta_i=e^{q_i-p_i}$ and plug into (\ref{eq:555}) and (\ref{eq:888}), we get
\begin{equation}
P_i(\mathcal{S}, \mathbf{p})=\frac{e^{q_i-p_i}}{\sum_{i\in \mathcal{S}}e^{q_i-p_i}+1}
\end{equation} where  $\mathbf{p}=\{p_{i}\mid i\in \Omega\}\in \mathbb{R}^{|\Omega|}$ is a price vector.
The expected revenue $g(\mathcal{S})$ of $\mathcal{S}$ is
\begin{equation}
g(\mathcal{S}, \mathbf{p})=\sum_{i\in \mathcal{S}} (p_i-c_i)P_i(\mathcal{S})
\end{equation}

The expected revenue of $S$ and price $\mathbf{p}$ can be calculated  as
\[f(S, \mathbf{p})=\sum_{i\in \mathcal{S}}\Pr[\mathcal{S}_{\leq i}\mid S] g(\mathcal{S}_{\leq i},\mathbf{p})\]

Our objective is to jointly decide a sequence of products and their prices to maximize the expected revenue subject to a cardinality constraint. We next present the formal definition of the second problem $\textbf{P.B}$.
 \begin{center}
\framebox[0.45\textwidth][c]{
\enspace
\begin{minipage}[t]{0.45\textwidth}
\small
$\textbf{P.B}$
$\max f(S, \mathbf{p})$\\
\textbf{subject to:} $|S|\leq B$;
\end{minipage}
}
\end{center}
\vspace{0.1in}

\section{Revenue maximizing product selection and sequencing}
We first study the case when $\alpha_i$ and $\beta_i$ are fixed for all products $i\in \Omega$. Our objective is to identify the best sequence of products that maximizes the expected revenue subject to a cardinality constraint. 
 Before presenting our solution to $\textbf{P.A}$,  we show that given any optimal solution $O$ to $\textbf{P.A}$, we can safely remove  those products whose reachability is sufficiently small from $O$ such that it does not affect the expected revenue of $O$ much.
\begin{lemma}
\label{lem:1}
For any $\rho\in[0,1]$, there is a solution $Q$ of expected revenue at least
\[f(Q)\geq (1-\rho)f(O)\] such that 
$|Q|\leq B$ and $\forall i\in Q: \Theta_i^{Q} \geq \rho$.
\end{lemma}
 \emph{Proof:}  Let $O[t]$ denote the $t$-th product in $O$. Assume $O[k]$ is the last question in $O$ whose reachability is no smaller than $\rho$, e.g.,  $k=\arg\max_{t} (\Theta_{O[t]}^{O}\geq \rho)$. Recall that we use $O_{> k}$ (resp. $O_{\leq k}$) to denote the sequence of questions scheduled after (resp. no later than) slot $k$. Therefore the reachability of every question in $O_{\leq k}$ is no smaller than $\rho$.

We first show that $\rho f(O_{> k})\geq  f(O)-f(O_{\leq k})$. Let $e_S$ denote the event that a consumer browses all and only products in $S$. We define $\Lambda_1=\{O_{\leq t} \mid t\leq k\}$ and  $\Lambda_2=\{ O_{\leq t}  \mid t>k\}$.
\begin{eqnarray}
f(O)&=& \sum_{A\in \Lambda_1} \Pr[e_A] g(\mathcal{A})+ \Pr[e_{O_{\leq k}}] g(\mathcal{O}_{\leq k})+ \sum_{A\in \Lambda_2}\Pr[e_A]g(\mathcal{A})~\nonumber\\
&\leq& \sum_{A\in \Lambda_1} \Pr[e_A] g(\mathcal{A})+ \Pr[e_{O_{\leq k}}] g(\mathcal{O}_{\leq k})+ \sum_{A\in \Lambda_2}\Pr[e_A] (g(\mathcal{A})-g(\mathcal{O}_{\leq k})+g(\mathcal{O}_{\leq k}) )\label{eq:6565}\\
&=& \sum_{A\in \Lambda_1} \Pr[e_A] g(\mathcal{A})+ \Pr[e_{O_{\leq k}}] g(\mathcal{O}_{\leq k})+ \sum_{A\in \Lambda_2}\Pr[e_A]g(\mathcal{O}_{\leq k})+  \sum_{A\in \Lambda_2}\Pr[e_A] (g(\mathcal{A})-g(\mathcal{O}_{\leq k}) )~\nonumber\\
&\leq&   f(O_{\leq k})+  \sum_{A\in \Lambda_2}\Pr[e_A] (g(\mathcal{A})-g(\mathcal{O}_{\leq k}) ) \label{eq:9090}\\
&\leq&  f(O_{\leq k})+  \sum_{A\in \Lambda_2}\Pr[e_A ] g(\mathcal{A}\setminus \mathcal{O}_{\leq k})  \label{eq:90901}\\
&=&  f(O_{\leq k})+ \Theta_{O[k]}^{O} \theta_{O[k]} f(O_{> k})\\
&\leq&  f(O_{\leq k})+ \rho f(O_{> k}) \label{eq:dun}
\end{eqnarray}
Inequality (\ref{eq:9090}) is due to  $\sum_{A\in \Lambda_1} \Pr[e_A] g(\mathcal{A})+ \Pr[e_{O_{\leq k}}] g(\mathcal{O}_{\leq k})+ \sum_{A\in \Lambda_2}\Pr[e_A]g(\mathcal{O}_{\leq k}) = f(O_{\leq k})$. Inequality (\ref{eq:dun}) is due to  $k=\arg\max_{t} (\Theta_{O[t]}^{O}\geq \rho)$.
It remained to prove inequality (\ref{eq:90901}), e.g.,  $g(\mathcal{A})-g(\mathcal{O}_{\leq k})\leq g(\mathcal{A}\setminus \mathcal{O}_{\leq k})$ for any $\mathcal{A} \in \Lambda_2$.
\begin{eqnarray}
g(\mathcal{A})-g(\mathcal{O}_{\leq k})&=&\sum_{i\in \mathcal{A}}\frac{ \alpha_i\beta_i}{\sum_{i\in \mathcal{A}}\beta_i+1}-\sum_{i\in \mathcal{O}_{\leq k}} \frac{\alpha_i\beta_i}{\sum_{i\in \mathcal{O}_{\leq k}}\beta_i+1}\\
&=& \sum_{i\in \mathcal{A}\cap \mathcal{O}_{\leq k}}\alpha_i(\frac{\beta_i}{\sum_{i\in \mathcal{A}}\beta_i+1}-\frac{\beta_i}{\sum_{i\in \mathcal{O}_{\leq k}}\beta_i+1}) \\
&&+ \sum_{i\in\mathcal{A}\setminus \mathcal{O}_{\leq k}} \frac{\alpha_i\beta_i}{\sum_{i\in \mathcal{A}}\beta_i+1}\\
&\leq& \sum_{i\in\mathcal{A}\setminus \mathcal{O}_{\leq k}} \frac{\alpha_i\beta_i}{\sum_{i\in \mathcal{A}}\beta_i+1}\\
&\leq& \sum_{i\in\mathcal{A}\setminus \mathcal{O}_{\leq k}} \frac{\alpha_i\beta_i}{\sum_{i\in \mathcal{A}\setminus \mathcal{O}_{\leq k}}\beta_i+1}=g(\mathcal{A}\setminus \mathcal{O}_{\leq k})
\end{eqnarray}

This finishes the proof of $\rho f(O_{> k})\geq  f(O)-f(O_{\leq k})$. Due to $O$ is the optimal solution to $\textbf{P.A}$, we have $f(O_{> k})\leq f(O)$. It follows that $(1-\rho)f(O)\leq f(O_{\leq k})$. Because $|O_{\leq k}|\leq B$ and all products in $O_{\leq k}$ can be browsed with probability at least $\rho$, $O_{\leq k}$ is one such sequence that satisfies all conditions in Lemma \ref{lem:1}. $\Box$

\begin{lemma}
\label{lem:2}
For any $\rho\in[0,1]$, there is a sequence $R$ with $|R|\leq b$ and  $\forall i\in R: \Theta_i^{R} \geq \rho$ such that
\[g(\mathcal{R})\geq (1-\rho)f(O)\] 
\end{lemma}
\emph{Proof:} Based on Lemma \ref{lem:1}, for any $\rho\in[0,1]$, there is a solution $Q$ of expected profit at least $f(Q)\geq (1-\rho)f(O)$.
Let $Q[t]$ denote the $t$-th product in $Q$. Assume that $|Q|=k$, we have
\[f(Q)=\sum_{t\in[1, k-1]}\Theta_{Q[t]}^{Q}(1-c_{Q[t]})g(\mathcal{Q}_{\leq t})+
\Theta_{Q[k]}^{Q}g(\mathcal{Q}_{\leq k})\geq (1-\rho)f(O)\]
The equality is due to the definition of $f(Q)$.
Because $\sum_{t\in[1,k-1]}\Theta_{Q[t]}^{Q}(1-c_{Q[t]})+ \Theta_{Q[k]}^{Q}=1$, we have
\[\max_{t\in[1,k]} g(\mathcal{Q}_{\leq t}) \geq (1-\rho)f(O)\]
Because $\max_{t\in[1,k]} g(\mathcal{Q}_{\leq t})$ is a subsequence of $Q$, we have $|\max_{t\in[1,k]} g(\mathcal{Q}_{\leq t})|\leq B$ and every product in $\max_{t\in[1,k]} g(\mathcal{Q}_{\leq t})$ has readability at least $\rho$. Therefore, $\max_{t\in[1,k]} g(\mathcal{Q}_{\leq t})$ is one such sequence that satisfies all conditions in Lemma \ref{lem:2}. $\Box$

 Based on Lemma \ref{lem:2}, in order to obtain a near-optimal solution, it suffice to consider those products whose reachability is sufficiently high. This motivates us to introduce a new problem $\textbf{P.A.1}$. The goal of  $\textbf{P.A.1}$ is to find a sequence of products that maximizes the expected revenue while ignoring those products whose reachability is sufficiently small. The solution to $\textbf{P.A.1}$ is composed of two parts: a set of products $\mathcal{S}$ and a single product $y\in \Omega$, where $y$ is displayed after $\mathcal{S}$. The reason we separate $y$ from other products in $\mathcal{S}$ is that $y$ is scheduled at the last slot, thus there is no restriction on $y$'s  continuation probability. Constraint (C1) ensures that every product can be viewed with probability at least $\rho$, and constraint (C2) ensures that the size of our solution is upper bounded by $B$.

   \begin{center}
\framebox[0.45\textwidth][c]{
\enspace
\begin{minipage}[t]{0.45\textwidth}
\small
$\textbf{P.A.1}$
\emph{Maximize$_{y, \mathcal{S} \subseteq \Omega \setminus \{y\}}$ $g(\mathcal{S}\cup y)$}\\
\textbf{subject to:}
\begin{equation*}
\begin{cases}
-\sum_{i\in \mathcal{S}} \log (\theta_i) \leq -\log \rho\quad \mbox{(C1)}\\
|\mathcal{S}| < B \quad \mbox{(C2)}
\end{cases}
\end{equation*}
\end{minipage}
}
\end{center}
\vspace{0.1in}
 We next present our algorithm Algorithm 1.

\textbf{Description of Algorithm 1.}
\begin{enumerate}
\item We first propose a fully  polynomial-time approximation scheme  (FPTAS) for $\textbf{P.A.1}$.
\item After solving  $\textbf{P.A.1}$ (approximately) and obtaining a solution $(\mathcal{S}^{\mathrm{alg_1}},y^{\mathrm{alg_1}})$, we build the final solution by first displaying $S^{\mathrm{alg_1}}$ (an arbitrary sequence of $\mathcal{S}^{\mathrm{alg_1}}$) and then displaying $y^{\mathrm{alg_1}}$.
\end{enumerate}

 In the rest of this section, we first present a FPTAS for $\textbf{P.A.1}$ and then analyze the performance of Algorithm 1.

 \subsection{FPTAS for $\textbf{P.A.1}$}

 Our basic idea, inspired by \cite{kempe2008cascade}, is to enumerate the last product $y\in\Omega$ in the optimal solution to $\textbf{P.A.1}$, for each fixed $y$, we present a FPTAS for $\textbf{P.A.1}$. At last, we choose the solution with the largest expected revenue as the final solution to $\textbf{P.A.1}$. Note that given a fixed $y$,  $\textbf{P.A.1}$ reduces to an assortment optimization problem subject to one capacity constraint and one cardinality constraint. \citep{desir2014near} develops a FPTAS for the assortment optimization problem subject to one capacity constraint, we extend their solution and provide a FPTAS, inspired by \cite{caprara2000approximation}, to  $\textbf{P.A.1}$ when $y$ is fixed, e.g., we provide the first FPTAS for the assortment optimization problem subject to one capacity constraint and one cardinality constraint.

Note that when $y$ is fixed, we only consider those products in $\Omega \setminus \{y\}$. For ease of presentation, we relabel all products in $\Omega \setminus \{y\}$ such that $\Omega \setminus \{y\}=[N-1]$. We first introduce some notations. Let $\alpha_{\min}=\min_{i\in \Omega} \alpha_i$ be the minimum revenue of a single product and  $\alpha_{\max}=\max_{i\in \Omega} \alpha_i$ be the maximum revenue of a single product. Let $\beta_{\min}=\min_{i\in \Omega} \beta_i$ and $\beta_{\max}=\max_{i\in \Omega} \beta_i$. Let $\gamma_i=\alpha_i\beta_i$, $\gamma_{\min}=\min_{i\in \Omega} \gamma_i$, and $\gamma_{\max}=\max_{i\in \Omega} \gamma_i$.

For a given $\epsilon>0$, we build the following group of guesses.
\[I=\{\gamma_{\min}(1+\epsilon)^a\mid a\leq \ln\frac{N\gamma_{\max}}{\epsilon\gamma_{\min}}\}, J=\{\beta_{\min}(1+\epsilon)^b\mid b\leq \ln\frac{N\beta_{\max}}{\epsilon\beta_{\min}}\}\]

For a given guess $\gamma_{\min}(1+\epsilon)^a\in I$ and $\beta_{\min}(1+\epsilon)^b\in J$, we discretize the values of $\gamma_i$ and $\beta_i$ as follows,
\[\tilde{\gamma_i}=\lceil\frac{\gamma_i}{\gamma_{\min}(1+\epsilon)^a\epsilon/B}\rceil, \tilde{\beta_i}=\lfloor\frac{\beta_i}{\beta_{\min}(1+\epsilon)^b\epsilon/B}\rfloor\]

We use $\omega_i$ to denote  $-\log (\theta_i)$ for all $i\in \Omega \setminus \{y\}$. Denote by function $h(j, u, v, l)$ for $j\in[N], u\in[\lceil N^2/\epsilon\rceil], v\in[\lceil N^2/\epsilon\rceil], l\in[B]$ the optimal solution value of the following problem:
\[h(j, u, v, l):= \min\{\sum_{i=1}^{j}\omega_ix_i: \sum_{i=1}^{j}\tilde{\gamma_i}x_i=u, \sum_{i=1}^{j}\tilde{\beta_i}x_i=v, \sum_{i=1}^{j}x_i=l, x_i\in\{0,1\}, l\in[B]\}\]
We set the initial values as follows: we first set $h(j, u, v, l)=+\infty$ for $ i=0, u\in[\lceil N^2/\epsilon \rceil], v\in[\lceil N^2/\epsilon\rceil], l\in[B]$, and then set $h(0, 0, 0, 0)=0$.

Then we fill up the dynamic program table using the following recurrence function.
\begin{equation*}
h(j, u, v, l)=
\begin{cases}
h(j-1, u, v, l)\quad &\mbox{if $u<\gamma_j$ or $v<\beta_j$};\\
 \min
\begin{cases}
h(j-1, u, v, l)\\
h(j-1, u-\gamma_j, v-\beta_j, l-1)+\omega_j
\end{cases}\quad &\mbox{otherwise}.
\end{cases}
\end{equation*}

After filling up the dynamic programming table, we go through all entries with $h(j, u, v, l)\leq -\log \rho$, and return the solution with the largest expected revenue.  We repeat the above process for every guess in $I\times J$, then return the one with the largest expected revenue as the final solution to $\textbf{P.A.1}$. Denote the returned solution by $(\mathcal{S}^{\mathrm{alg_1}}, y^{\mathrm{alg_1}})$. We next prove that the above solution achieves $\frac{1-\epsilon(1+\epsilon)}{1+\epsilon(1+\epsilon)}$ approximation ratio for $\textbf{P.A.1}$.

\begin{lemma}
\label{lem:14}
Let $(\mathcal{S}^*, y^*)$ denote the optimal solution to $\textbf{P.A.1}$. For any $\epsilon>0$,
\[g(\mathcal{S}^{\mathrm{alg_1}}\cup y^{\mathrm{alg_1}})\geq \frac{1-\epsilon(1+\epsilon)}{1+\epsilon(1+\epsilon)} g(\mathcal{S}^*\cup y^*)\]
\end{lemma}
\emph{Proof:} Assume  $\gamma_{\min}(1+\epsilon)^a \leq \sum_{i\in \mathcal{S}^*}\gamma_i \leq \gamma_{\min}(1+\epsilon)^{a+1}$ and $\beta_{\min}(1+\epsilon)^b \leq \sum_{i\in \mathcal{S}^*}\beta_i \leq \beta_{\min}(1+\epsilon)^{b+1}$. Recall that to obtain a FPTAS for $\textbf{P.A.1}$, we enumerate the last product $o\in\Omega$ and solve the dynamic program for each $y$ and each guess in $I\times J$. Consider the case when $(y^*,\gamma_{\min}(1+\epsilon)^{a+1}\in I,\beta_{\min}(1+\epsilon)^{b+1}\in J)$ is enumerated, let $u^*=\sum_{i\in \mathcal{S}^*}\tilde{\gamma_i}$ and $v^*=\sum_{i\in \mathcal{S}^*}\tilde{\beta_i}$ denote the summation of the scaled values. It is clear that $h(N, u^*, v^*, |\mathcal{S}^*|)\leq -\log\rho$, let  $S'$ denote the solution stored in $h(N, u^*, v^*, |\mathcal{S}^*|)$.
We first give a lower bound on  $\sum_{z\in S'} \gamma_z$,
\begin{eqnarray}
\sum_{z\in S'} \gamma_z &\geq& \sum_{z\in S'} \tilde{\gamma_z}\epsilon\gamma_{\min}(1+\epsilon)^{a+1}/B-\epsilon\gamma_{\min}(1+\epsilon)^{a+1}\\
&=& u^*\epsilon\gamma_{\min}(1+\epsilon)^{a+1}/B -\epsilon\gamma_{\min}(1+\epsilon)^{a+1}\\
&\geq& u^*\epsilon\gamma_{\min}(1+\epsilon)^{a+1}/B-\epsilon(1+\epsilon)\sum_{i\in \mathcal{S}^*}\gamma_i\\
&\geq& \sum_{i\in \mathcal{S}^*}\gamma_i-\epsilon(1+\epsilon)\sum_{i\in \mathcal{S}^*}\gamma_i\\
&=&(1-\epsilon(1+\epsilon))\sum_{i\in \mathcal{S}^*}\gamma_i
 \end{eqnarray} where the last inequality is due to $\tilde{\gamma_i}\geq \frac{\gamma_i}{\gamma_{\min}(1+\epsilon)^{a+1}\epsilon/n}$ for all $i\in \Omega$.

Then we give an upper bound on $\sum_{z\in S'} \beta_z$,
\begin{eqnarray}\sum_{z\in S'} \beta_z
&\leq& \sum_{z\in S'} \tilde{\beta_z}\epsilon\beta_{\min}(1+\epsilon)^{b+1}/B+\epsilon\beta_{\min}(1+\epsilon)^{b+1} \\
&=& v^*\epsilon\beta_{\min}(1+\epsilon)^{b+1}/B+\epsilon\beta_{\min}(1+\epsilon)^{b+1}\\
&\leq& v^*\epsilon\beta_{\min}(1+\epsilon)^{b+1}/B+\epsilon(1+\epsilon)\sum_{i\in \mathcal{S}^*}\beta_i \\
&\leq& \sum_{i\in \mathcal{S}^*}\beta_i+\epsilon(1+\epsilon)\sum_{i\in \mathcal{S}^*}\beta_i\\
&=&(1+\epsilon(1+\epsilon))\sum_{i\in \mathcal{S}^*}\beta_i
 \end{eqnarray} where the last inequality is due to $\tilde{\beta_i}\leq  \frac{\beta_i}{\beta_{\min}(1+\epsilon)^{a+1}\epsilon/n}$ for all $i\in \Omega$.

It follows that
\begin{eqnarray} g(\mathcal{S}^{\mathrm{alg_1}}\cup y^{\mathrm{alg_1}})&\geq& g(\mathcal{S}'\cup y^*)\\
&=&\frac{\sum_{z\in S'} \gamma_z+\gamma_{y^*}}{\sum_{z\in S'} \beta_z+\beta_{y^*}+1}\\
&\geq& \frac{(1-\epsilon(1+\epsilon))\sum_{i\in \mathcal{S}^*}\gamma_i+\gamma_{y^*}}{(1+\epsilon(1+\epsilon))\sum_{i\in \mathcal{S}^*}\beta_i+\beta_{y^*}+1}\\
&\geq&\frac{1-\epsilon(1+\epsilon)}{1+\epsilon(1+\epsilon)}g(\mathcal{S}^*\cup y^*)
\end{eqnarray} $\Box$

\subsection{Performance analysis of  Algorithm 1}
We next analyze the performance bound of Algorithm 1. Recall that after obtaining $(\mathcal{S}^{\mathrm{alg_1}},y^{\mathrm{alg_1}})$ from the previous stage, Algorithm 1 returns $S^{\mathrm{alg_1}}\oplus y^{\mathrm{alg_1}}$ as the final solution where $S^{\mathrm{alg_1}}$ is an arbitrary sequence of $\mathcal{S}^{\mathrm{alg_1}}$. We next prove that  for  any $\epsilon>0$ and $\rho\in[0,1]$, Algorithm 1 achieves $\frac{1-\epsilon(1+\epsilon)}{1+\epsilon(1+\epsilon)}\rho(1-\rho)$ approximation ratio.

\begin{theorem}For any $\epsilon>0$ and $\rho\in[0,1]$, \[f(S^{\mathrm{alg_1}}\oplus y^{\mathrm{alg_1}})\geq  \frac{1-\epsilon(1+\epsilon)}{1+\epsilon(1+\epsilon)}\rho(1-\rho)f(O)\]
\end{theorem}
\emph{Proof:} Due to constraint (C2), a consumer views all product in $S^{\mathrm{alg_1}}\oplus y^{\mathrm{alg_1}}$ with probability at least $\rho$, then we have
\begin{equation}
f(S^{\mathrm{alg_1}}\oplus y^{\mathrm{alg_1}})\geq \rho g(\mathcal{S}^{\mathrm{alg_1}}\cup y^{\mathrm{alg_1}})\label{eq:3232}
 \end{equation}
Based on Lemma \ref{lem:14}, we have \begin{equation}\label{eq:32323}
 g(\mathcal{S}^{\mathrm{alg_1}}\cup y^{\mathrm{alg_1}})\geq \frac{1-\epsilon(1+\epsilon)}{1+\epsilon(1+\epsilon)} g(\mathcal{S}^*\cup y^*)
\end{equation}
(\ref{eq:3232}) and (\ref{eq:32323}) imply that
 \begin{equation}\label{eq:3232321}
 f(S^{\mathrm{alg_1}}\oplus y^{\mathrm{alg_1}})\geq \rho \frac{1-\epsilon(1+\epsilon)}{1+\epsilon(1+\epsilon)} g(\mathcal{S}^*\cup y^*)  \end{equation}
 In Lemma \ref{lem:13} (whose proof we defer until later), we prove that
 \begin{equation}\label{eq:3232324}g(\mathcal{S}^*\cup y^*)\geq  (1-\rho)f(O)
  \end{equation}
(\ref{eq:3232321}) and (\ref{eq:3232324}) imply that $f(S^{\mathrm{alg_1}}\oplus y^{\mathrm{alg_1}})\geq  \frac{1-\epsilon(1+\epsilon)}{1+\epsilon(1+\epsilon)}\rho(1-\rho)f(O)$. $\Box$

We next focus on proving Lemma \ref{lem:13}.

\begin{lemma}
\label{lem:13}
For any $\rho\in[0,1]$,
\[g(\mathcal{S}^*\cup y^*)\geq  (1-\rho)f(O)\]
\end{lemma}
\emph{Proof:} Recall that in Lemma \ref{lem:2}, we prove that there exists a sequence $R$ with $|R|\leq B$ and  $\forall i\in R: \Theta_i^{R} \geq \rho$ such that
$g(\mathcal{R})\geq (1-\rho)f(O)$. Let $s$ denote the last product in $R$, it is easy to verify that $(\mathcal{R}\setminus s, s)$ is a feasible solution to \textbf{P.A.1}. As a result, $g(\mathcal{S}^*\cup y^*)\geq g(\mathcal{R})\geq (1-\rho)f(O)$. $\Box$


\section{Joint product selection, sequencing, and pricing}
We next study the case when the price of each product is also a decision variable. 
As described in Section \ref{sec:111}, assume each product $i\in \Omega$ is associated with a fixed quality $q_i$, an adjustable price $p_i$, and a fixed cost $c_i$, we set $\alpha_i=p_i-c_i$ and  $\beta_i=e^{q_i-p_i}$. Our objective is to find a solution $(S, \mathbf{p})$ to $\textbf{P.B}$, where $S$ is a sequence of products  and  $\mathbf{p}=\{p_i\mid i\in \Omega\}$ is the corresponding pricing vector. Let $(O,\mathbf{p}^{OPT})$ denote the optimal solution to $\textbf{P.B}$. Based on similar proofs of Lemma \ref{lem:1} and \ref{lem:2}, we have the following two lemmas.

\begin{lemma}
\label{lem:7}
For any $\rho\in[0,1]$, there is a sequence $Q$ with $|Q|\leq B$ and $\forall i\in Q: \Theta_i^{Q} \geq \rho$ such that
\[f(Q,\mathbf{p}^{OPT})\geq (1-\rho)f(O,\mathbf{p}^{OPT})\]
\end{lemma}
\emph{Proof Sketch:} We fix the price $\mathbf{p}^{OPT}$, then apply the proof of  Lemma \ref{lem:1} to complete the proof.
\begin{lemma}
\label{lem:8}
For any $\rho\in[0,1]$, there is a sequence $R$ with $|R|\leq B$ and  $\forall i\in R: \Theta_i^{R} \geq \rho$ and price $\mathbf{p}$ such that
\[g(\mathcal{R},\mathbf{p}^{OPT})\geq (1-\rho)f(O,\mathbf{p}^{OPT})\] 
\end{lemma}
\emph{Proof Sketch:} We fix the price $\mathbf{p}^{OPT}$, then apply the proof of  Lemma \ref{lem:2} to complete the proof.

We next introduce a new problem $\textbf{P.B.1}$ whose goal is to jointly decide a consideration set and a pricing to maximize the expected revenue, while ignoring those products whose reachability is sufficiently small. Similar to $\textbf{P.A.1}$, we still use $y$ to denote the last product in a sequence, (C1) ensures that all products in a feasible solution can be viewed with probability at least $\rho$, and (C2) ensures that the cardinality of the solution is bounded by $B$. We next describe the design of our algorithm (Algorithm 2).
   \begin{center}
\framebox[0.45\textwidth][c]{
\enspace
\begin{minipage}[t]{0.45\textwidth}
\small
$\textbf{P.B.1}$
\emph{Maximize$_{y, \mathcal{S} \subseteq \Omega \setminus \{y\}, \mathbf{p}}$ $g(\mathcal{S}\cup y, \mathbf{p})$}\\
\textbf{subject to:} (C1) and (C2)
\end{minipage}
}
\end{center}

\textbf{Description of Algorithm 2.}
\begin{enumerate}
\item We first propose a FPTAS for $\textbf{P.B.1}$.
\item After solving  $\textbf{P.B.1}$ (approximately) and obtaining a $(1-\epsilon)$-approximate solution $(\mathcal{S}^{\mathrm{alg_2}},y^{\mathrm{alg_2}},\mathbf{p}^{\mathrm{alg_2}})$, we build the final solution by first displaying $S^{\mathrm{alg_2}}$ (an arbitrary sequence of $\mathcal{S}^{\mathrm{alg_1}}$) and then displaying $y^{\mathrm{alg_2}}$ using price $\mathbf{p}^{\mathrm{alg_2}}$.
\end{enumerate}

In the rest of this section, we first present a FPTAS for $\textbf{P.B.1}$ and then analyze the performance bound of Algorithm 2.

\subsection{A FPTAS for $\textbf{P.B.1}$}


To facilitate our study, we first introduce a well-known result in the field of assortment optimization.
\begin{lemma}
\label{lem:17}\cite{hopp2005product}
Given any consideration set $\mathcal{S}$, the maximum revenue $\max_{\mathbf{p}}g(\mathcal{S},\mathbf{p})$ is achieved at $p_i=W(\sum_{i\in \mathcal{S}}e^{q_i-c_i-1})+c_i+1$ for all products $i\in \Omega$ where $W(z)$ is the solution to $x e^x=z$. Moreover, the value of the maximum revenue is $\max_{\mathbf{p}}g(\mathcal{S},\mathbf{p})=W(\sum_{i\in \mathcal{S}}e^{q_i-c_i-1})$.
\end{lemma}

Based on Lemma \ref{lem:17}, we obtain a closed form solution of the maximum revenue for any consideration set. This enables us to remove the decision variable $\mathbf{p}$ from $\textbf{P.B.1}$ to obtain an equivalent, but simplified, formulation in $\textbf{P.B.2}$. In particular, we replace the objective function $g(\mathcal{S}\cup y, \mathbf{p})$ in $\textbf{P.B.1}$ by $W(\sum_{i\in \mathcal{S}}v_i+v_y)$ in $\textbf{P.B.2}$. 

  \begin{center}
\framebox[0.55\textwidth][c]{
\enspace
\begin{minipage}[t]{0.55\textwidth}
\small
$\textbf{P.B.2}$
\emph{Maximize$_{y, \mathcal{S} \subseteq \Omega \setminus \{y\}}$ $W(\sum_{i\in \mathcal{S}}v_i+v_y)$}\\
\textbf{subject to:} (C1) and (C2)
\end{minipage}
}
\end{center}
Note that maximizing $W(\sum_{i\in \mathcal{S}}v_i+v_y)$ is equivalent to maximizing $\sum_{i\in \mathcal{S}}v_i+v_y$. We can further simplify $\textbf{P.B.2}$ by replacing the objective function $W(\sum_{i\in \mathcal{S}}v_i+v_y)$  by $\sum_{i\in \mathcal{S}}v_i+v_y$. We denote by $\textbf{P3.2}$ the new formulation of $\textbf{P.B.2}$.

  \begin{center}
\framebox[0.55\textwidth][c]{
\enspace
\begin{minipage}[t]{0.55\textwidth}
\small
$\textbf{P.B.3}$
\emph{Maximize$_{y, \mathcal{S} \subseteq \Omega \setminus \{y\}}$ $\sum_{i\in \mathcal{S}}v_i+v_y$}\\
\textbf{subject to:} (C1) and (C2)
\end{minipage}
}
\end{center}

 Now we are ready to present the solution to $\textbf{P.B.1}$. We first enumerate the last product $o\in \Omega$ and solve $\textbf{P.B.3}$ for each $y$.   The solution with the largest expected revenue is returned as the final solution to $\textbf{P.B.1}$.  Let $\textbf{P.B.3}(y)$ denote $\textbf{P.B.3}$ when the last product $y$ is given. It is easy to verify that for a fixed $y$, $\textbf{P.B.3}(y)$ is a classic knapsack problem subject to a capacity constraint and a cardinality constraint. Given that $\textbf{P.B.3}(y)$ admits an FPTAS \cite{caprara2000approximation}, $\textbf{P.B.3}$ also admits an FPTAS. A detailed proof is provided in Lemma \ref{lem:2222}.

  \begin{center}
\framebox[0.45\textwidth][c]{
\enspace
\begin{minipage}[t]{0.45\textwidth}
\small
$\textbf{P.B.3}(y)$
\emph{Maximize$_{\mathcal{S} \subseteq \Omega \setminus \{y\}}$ $\sum_{i\in \mathcal{S}}v_i$}\\
\textbf{subject to:} (C1) and (C2)
\end{minipage}
}
\end{center}

\begin{lemma}
\label{lem:2222}
If $\textbf{P.B.3}(y)$ admits an FPTAS for any fixed $y$,  $\textbf{P.B.1}$ also admits an FPTAS.
\end{lemma}
\emph{Proof:} Recall that to solve $\textbf{P.B.1}$, we solve $\textbf{P.B.3}(y)$ approximately for all $y \in \Omega$ and choose the best one as the final solution. Because  $\textbf{P.B.1}$ and  $\textbf{P.B.2}$ are equivalent, to prove this lemma, it is equivalent to proving that any $(1-\epsilon)$-approximate solution to $\textbf{P.B.3}(y)$ is also a $(1-\epsilon)$-approximate solution to $\textbf{P.B.2}$  for a fixed $y$ and any $\epsilon<1$. Given any $y$, assume there exists a $(1-\epsilon)$-approximate solution $\mathcal{D}$  to  $\textbf{P.B.3}(y)$, i.e.,  $\sum_{i\in \mathcal{D}}v_i\geq (1-\epsilon) \max_{\mathcal{S}}(\sum_{i\in \mathcal{S}}v_i)$ subject to all constraints in $\textbf{P.B.3}(y)$. It follows that $\sum_{i\in \mathcal{D}}v_i+v_y\geq (1-\epsilon) (\max_{\mathcal{S}}\sum_{i\in \mathcal{S}}v_i+v_y)$. It implies that $W(\sum_{i\in \mathcal{D}}v_i+v_y)\geq W((1-\epsilon) (\max_{\mathcal{S}}\sum_{i\in \mathcal{S}}v_i+v_y))$. Because $W$ is concave, we have  $W(\sum_{i\in \mathcal{D}}v_i+v_y)\geq (1-\epsilon)  W(\max_{\mathcal{S}}(\sum_{i\in \mathcal{S}}v_i+v_y))=(1-\epsilon)  \max_{\mathcal{S}} W(\sum_{i\in \mathcal{S}}v_i+v_y)$. $\Box$

\subsection{Performance analysis of Algorithm 2}
We next analyze the performance bound of Algorithm 2. We use  $(\mathcal{S}^*, y^*, \mathbf{p}^*)$ to denote the optimal solution to $\textbf{P.B.1}$.
\begin{theorem}
\label{thm:2}
For any $\epsilon<1$, we have \[f(\mathcal{S}^{\mathrm{alg_2}}\oplus y^{\mathrm{alg_2}}, \mathbf{p}^{\mathrm{alg_2}})\geq  (1-\epsilon)\rho (1-\rho)f(O,\mathbf{p}^{OPT})\]
\end{theorem}
\emph{Proof:} Recall that in Lemma \ref{lem:8}, we prove that there exists a sequence $R$ with $|R|\leq B$ and  $\forall i\in R: \Theta_i^{R} \geq \rho$ such that
$g(\mathcal{R},\mathbf{p}^{OPT})\geq (1-\rho)f(O, \mathbf{p}^{OPT})$. Let $s$ denote the last product in $R$, it is easy to verify that $(\mathcal{R}\setminus s, s, \mathbf{p}^{OPT})$ is a feasible solution to \textbf{P.B.1}. As a result,
\begin{eqnarray}
g(\mathcal{S}^*\cup y^*, \mathbf{p}^*) &\geq& g(\mathcal{R},\mathbf{p}^{OPT})\\
&\geq& (1-\rho)f(O, \mathbf{p}^{OPT}) \label{eq:1}
\end{eqnarray}

The first inequality is due to $(\mathcal{S}^*, y^*, \mathbf{p}^*)$ is the optimal solution to $\textbf{P.B.1}$.

It follows that
\begin{eqnarray}
f(\mathcal{S}^{\mathrm{alg_2}}\oplus y^{\mathrm{alg_2}}, \mathbf{p}^{\mathrm{alg_2}})
&\geq& \rho g(\mathcal{S}^{\mathrm{alg_2}}\cup y, \mathbf{p}^{\mathrm{alg_2}})\\
&\geq&  (1-\epsilon)\rho(1-\rho)f(O,\mathbf{p}^{OPT})
\end{eqnarray}
The first inequality is due to the following observation: Because $(\mathcal{S}^{\mathrm{alg_2}},y^{\mathrm{alg_2}})$  is a feasible solution to \textbf{P.A.1}, constraint (C2) ensures that all products in $S^{\mathrm{alg_2}}\oplus y^{\mathrm{alg_2}}$ can be browsed with probability at least $\rho$. Thus, $f(\mathcal{S}^{\mathrm{alg_2}}\oplus y^{\mathrm{alg_2}}, \mathbf{p}^{\mathrm{alg_2}})\geq \rho g(\mathcal{S}^{\mathrm{alg_2}}\cup y, \mathbf{p}^{\mathrm{alg_2}})$. The second inequality is due to the assumption that $\textbf{P.B.1}$ admits an FPTAS, i.e., $g(\mathcal{S}^{\mathrm{alg_2}}\cup y, \mathbf{p}^{\mathrm{alg_2}})\geq (1-\epsilon)g(\mathcal{S}^*\cup y^*, \mathbf{p}^*)$ for any $\epsilon<1$, and (\ref{eq:1}). $\Box$
\vspace{0.1in}
\section{Conclusion}
We study the product sequencing and pricing problem under the cascade browse model. In the first setting, we assume that both the revenue and MNL-based preference weight are fixed for all products, and focus on finding the best sequence of products subject to a cardinality constraint. In the second setting, we consider the joint sequencing and pricing problem. We develop approximate solutions to both settings. As a by product, we propose the first FPTAS for the assortment optimization problem subject to one capacity constraint and one cardinality constraint.
\bibliographystyle{model1a-num-names}
\bibliography{reference}

\begin{thebibliography}{10}
\expandafter\ifx\csname url\endcsname\relax
  \def\url#1{\texttt{#1}}\fi
\expandafter\ifx\csname urlprefix\endcsname\relax\def\urlprefix{URL }\fi
\expandafter\ifx\csname href\endcsname\relax
  \def\href#1#2{#2} \def\path#1{#1}\fi

\bibitem{aouad2019click}
A.~Aouad, J.~Feldman, D.~Segev, D.~Zhang, Click-based mnl: Algorithmic
  frameworks for modeling click data in assortment optimization, Available at
  SSRN 3340620.

\bibitem{craswell2008experimental}
N.~Craswell, O.~Zoeter, M.~Taylor, B.~Ramsey, An experimental comparison of
  click position-bias models, in: Proceedings of the 2008 international
  conference on web search and data mining, ACM, 2008, pp. 87--94.

\bibitem{li2015d}
G.~Li, P.~Rusmevichientong, H.~Topaloglu, The d-level nested logit model:
  Assortment and price optimization problems, Operations Research 63~(2) (2015)
  325--342.

\bibitem{davis2014assortment}
J.~M. Davis, G.~Gallego, H.~Topaloglu, Assortment optimization under variants
  of the nested logit model, Operations Research 62~(2) (2014) 250--273.

\bibitem{blanchet2016markov}
J.~Blanchet, G.~Gallego, V.~Goyal, A markov chain approximation to choice
  modeling, Operations Research 64~(4) (2016) 886--905.

\bibitem{farias2013nonparametric}
V.~F. Farias, S.~Jagabathula, D.~Shah, A nonparametric approach to modeling
  choice with limited data, Management science 59~(2) (2013) 305--322.

\bibitem{aouad2015assortment}
A.~Aouad, V.~F. Farias, R.~Levi, Assortment optimization under
  consider-then-choose choice models, Available at SSRN 2618823.

\bibitem{davis2013assortment}
J.~Davis, G.~Gallego, H.~Topaloglu, Assortment planning under the multinomial
  logit model with totally unimodular constraint structures, Work in Progress.

\bibitem{abeliuk2016assortment}
A.~Abeliuk, G.~Berbeglia, M.~Cebrian, P.~Van~Hentenryck, Assortment
  optimization under a multinomial logit model with position bias and social
  influence, 4OR 14~(1) (2016) 57--75.

\bibitem{aouad2015display}
A.~Aouad, D.~Segev, Display optimization for vertically differentiated
  locations under multinomial logit choice preferences, Available at SSRN
  2709652.

\bibitem{ferreira2019learning}
K.~Ferreira, S.~Parthasarathy, S.~Sekar, Learning to rank an assortment of
  products, Available at SSRN 3395992.

\bibitem{wang2017impact}
R.~Wang, O.~Sahin, The impact of consumer search cost on assortment planning
  and pricing, Management Science 64~(8) (2017) 3649--3666.

\bibitem{anderson1992discrete}
S.~P. Anderson, A.~De~Palma, J.-F. Thisse, Discrete choice theory of product
  differentiation, MIT press, 1992.

\bibitem{song2007demand}
J.-S. Song, Z.~Xue, Demand management and inventory control for substitutable
  products, W orking paper.

\bibitem{hopp2005product}
W.~J. Hopp, X.~Xu, Product line selection and pricing with modularity in
  design, Manufacturing \& Service Operations Management 7~(3) (2005) 172--187.

\bibitem{najafi2019multi}
S.~Najafi, I.~Duenyas, S.~Jasin, J.~Uichanco, Multi-product dynamic pricing
  with limited inventories under cascade click model, Available at SSRN
  3362921.

\bibitem{kempe2008cascade}
D.~Kempe, M.~Mahdian, A cascade model for externalities in sponsored search,
  in: International Workshop on Internet and Network Economics, Springer, 2008,
  pp. 585--596.

\bibitem{desir2014near}
A.~D{\'e}sir, V.~Goyal, J.~Zhang, Near-optimal algorithms for capacity
  constrained assortment optimization, Available at SSRN 2543309.

\bibitem{caprara2000approximation}
A.~Caprara, H.~Kellerer, U.~Pferschy, D.~Pisinger, Approximation algorithms for
  knapsack problems with cardinality constraints, European Journal of
  Operational Research 123~(2) (2000) 333--345.

\end{thebibliography}

\end{document}